\documentclass[prb,reprint,twocolumn,showpacs,superscriptaddress,floatfix,aps]{revtex4-2}

\usepackage{amssymb,amsmath,latexsym,bm,dsfont,graphicx,braket}
\usepackage{txfonts}
\usepackage{kotex} 
\makeatletter\def\@pacs@name{DOI: }\makeatother 

\renewcommand{\vec}[1]{{\bm{\mathrm{#1}}}}
\newcommand{\vhat}[1]{\hat{\bm{\mathrm{#1}}}}
\let\epsilon\varepsilon

\DeclareMathOperator{\sgn}{sgn}

\usepackage{xcolor}

\usepackage[colorlinks=true, citecolor=blue, linkcolor=blue, urlcolor=blue]{hyperref}

\newcommand{\Letter}{paper}

\newcommand{\EndMatter}{\quad}

\begin{document}
\title{Orbital-driven emergent transport in altermagnets}
\author{Junyeong Choi}%
\author{Kyoung-Whan Kim}%
\email{kwkim@yonsei.ac.kr}
\affiliation{Department of Physics, Yonsei University, Seoul 02792, Korea}%
\affiliation{Center for Quantum Dynamics of Angular Momentum, Pohang University of Science and Technology, Pohang 37673, Korea}%
\date{\today}

\begin{abstract}
Altermagnets have recently emerged as a promising platform for spintronics due to their unique magnetic symmetry. However, most studies have focused on spin degrees of freedom, leaving the dynamic role of orbital degrees of freedom largely unexplored. In this work, we extend the altermagnet Hamiltonian to include the orbital degree of freedom as a dynamical variable and derive the resulting emergent electromagnetic fields (EEMFs). This approach allows us to demonstrate emergent electric fields controllable via lattice anisotropy and the resulting orbital and magnetic multipole currents. Furthermore, we show that non-vanishing emergent electric fields can arise even in simplified spin and orbital textures, particularly in the presence of dynamic lattice distortion. This formalism is generalizable to high-order altermagnets beyond $d$-wave systems. 
\end{abstract}

\pacs{\quad 
}

\maketitle

\section{Introduction} In conventional antiferromagnets, the absence of net magnetization is associated with $\mathcal{PT}$ symmetry which enforces spin degeneracy and prohibits spin splitting in momentum space. Recently, altermagnets have been proposed as a distinct magnetic phase exhibiting pronounced momentum-dependent spin splitting~\cite{Krempasky2024} that arises from their crystal symmetries and persists even in the absence of both net magnetization and spin-orbit coupling~\cite{emerging,alterfirst,Beyond}. The anisotropy of the local crystal field, induced by the sublattice crystal symmetry, gives rise to the unusual orbital symmetry the of magnetic atoms in altermagnets, serving as the origin of the momentum-dependent spin splitting. Due to the substantial momentum-dependent spin splitting, altermagnets are expected to facilitate the efficient generation and manipulation of spin transport, establishing them as a promising materials for spintronic applications.

In addition to these momentum-space mechanisms, spin transport can also be driven by spatiotemporal dynamics of spin textures, such as moving skyrmions, via emergent electromagnetic fields (EEMFs). In strong ferromagnets, EEMFs generated through the $s$-$d$ exchange interaction, $\mathcal{H}_\text{FM}=\hbar^2\vec{k}^2/2m_e + J \vec{\sigma} \cdot \vec{m}(\vec{r},t)$, have been extensively studied~\cite{volovik1987linear,nastt,yamaneequation,Intrinsic}, leading to prominent phenomena such as the spin motive force~\cite{SMFmeasure,Faraday,Giant} and the topological Hall effect~\cite{MnSi,THEandBerry}. Extending this framework to altermagnets using the widely used Hamiltonian~\cite{roig2024minimal},
\begin{equation}
	\mathcal{H}_\text{AM}=\hbar^2\vec{k}^2/2m_e + J_s(\vec{k}) \vec{\sigma} \cdot \vec{n}(\vec{r},t),\label{0}
\end{equation}	
yields the EEMFs given by
\begin{align}
	\vec{E}_{\text{AM},s} &= \frac{\hbar}{2e} \sgn[sJ(\vec{k})] \sum_i [\vec{n} \cdot (\partial_t \vec{n} \times \partial_i \vec{n})] \vhat{e}_i, \label{1} \\
	\vec{B}_{\text{AM},s} &= -\frac{\hbar}{4e} \sgn[sJ(\vec{k})] \sum_i \varepsilon^{ijk}[\vec{n} \cdot (\partial_j \vec{n} \times \partial_k \vec{n})] \vhat{e}_i,	\label{2}
\end{align}
where $\vec{k}$ is the crystal momentum of electrons, $m_e$ is the electron mass, $e<0$ is the electron charge, $J_s(\vec{k})$ represents the momentum-dependent spin splitting in altermagnets, the $\vec{n}(\vec{r},t)$ denotes the N\'{e}el vector, which is the well-known order parameter of antiferromagnets, $s = \pm 1$ (or $\uparrow, \downarrow$) is the spin index denoting the conduction electron spin projection parallel or anti-parallel to $\vec{n}$, and $\vec{\sigma}$ represents the spin Pauli matrix. For $d$-wave altermagnets for example, $J_s(\vec{k})\propto k_xk_y$ or $(k_x^2-k_y^2)$, so the sign of the EEMFs depends not only on $s$, but also on the direction of $\mathbf{k}$. Because of this momentum dependence, relying solely on the spin degree of freedom is insufficient since the macroscopic EEMF contributions could cancel out upon integration over the Brillouin zone. Therefore, an explicit treatment of the orbital degree of freedom is required to accurately capture these emergent transport phenomena.

The generation of these EEMFs manifests as the spin motive force, which is the reciprocal process of spin-transfer torque (STT). While orbitronics has recently attracted considerable attention, investigations into related reciprocal processes such as orbital pumping~\cite{Han2025,Hayashi2024,Go2025} remain restricted to heterostructures. This limitation arises because the orbital angular momentum in condensed matters is typically quenched in equilibrium or induced by spin-orbit coupling, making it difficult to distinguis it from spin measurements. However, altermagnets fundamentally overcome this issue by allowing the spin and orbital degrees of freedom to act independently without relying on strong spin-orbit coupling, and by stabilizing orbital-driven multipoles~\cite{Han2025Octupole,Ko2025,Bhowal2024,MultipoleTani} as their equilibrium order parameters. Therefore, constructing altermagnet Hamiltonians that accounts for dynamic orbital degrees of freedom is a necessary step for a correct understanding of the responses arising from altermagnets, and is expected to reveal new physical phenomena beyond standard STT and spin motive force.

In this \Letter, we extend the Hamiltonian of general altermagnets to include the orbital degree of freedom. By applying a gauge transformation~\cite{Tatara2019}, we derive the band-dependent EEMFs. Most notably, we demonstrate the existence of EEMFs that originate from the orbital degree of freedom, which have not been previously reported. To relate the derived EEMFs to experimental observables, we consider the simplest $d$-wave altermagnet to 
calculate the linear response of charge, spin, orbital, and magnetic octupole transport driven by EEMFs under the Drude model. We reveal that the magnetic octupole current driven by the emergent electric field is directly manipulated by the intrinsic band-anisotropy of altermagnets, yielding a novel contribution fundamentally absent in conventional antiferromagnets. Furthermore, 
accounting for the orbital degree of freedom allows us to consider dynamic lattice distortions, which turns out to yield non-zero EEMFs under conditions feasible for direct experimental verification.

\section{EEMFs with the orbital degree of freedom}
\begin{figure}[t]
	\centering
	\includegraphics[width=1.0\linewidth]{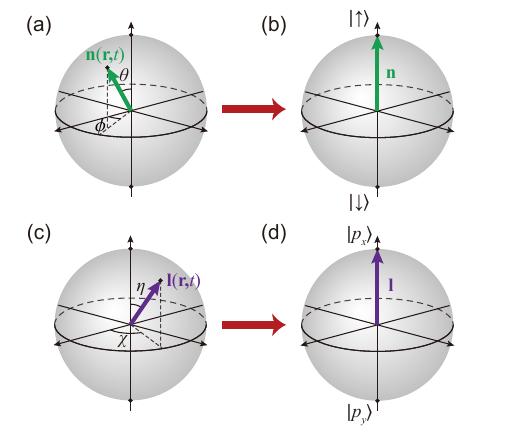}
	\caption{Schematic of (a) the N\'{e}el vector $\vec{n}$ and (c) the orbital N\'{e}el vector $\vec{l}$ on the Bloch sphere, treated as a pseudospins. (b) and (d) indicate the fixed N\'{e}el vector and orbital N\'{e}el vector via a gauge transformation. In this figure, the orbital basis $\{\ket{p_x}, \ket{p_y} \}$ is used as an example.}
	\label{fig:2}
\end{figure}

To consider the dynamic orbital degree of freedom, it is necessary to start with a Hamiltonian incorporating both the orbital degree of freedom and the band anisotropy. For altermagnets, there are two active orbitals, such as $\{\ket{p_x}, \ket{p_y} \}$ or $\{ \ket{d_{3z^2-r^2}}, \ket{d_{xy}} \}$, associated with the spin degree of freedom. We adopt the corresponding pseudospin basis spanned by the orbitals. 
To capture the momentum-dependent spin splitting of altermagnets originating from the anisotropic orbitals, the band anisotropy is considered, by decomposing the dispersion relation into an isotropic term, $\varepsilon_0(\vec{k})$, and an anisotropic term, $\varepsilon_{\text{ani}}(\vec{k})$ 
in the spin-dependent Hamiltonian. However, the conventional descriptions~\cite{Berry,volovik1987linear} such as Eqs.~\eqref{1} and~\eqref{2} applied to the spin-dependent Hamiltonian [Figs.~\ref{fig:2}(a) and (b)] take into account the spin N\'{e}el vector $\vec{n}(\vec{r},t)$ exclusively, and thus fail to capture the orbital contribution to the EEMFs. 

To overcome this limitation, we introduce the orbital N\'{e}el vector $\vec{l}(\vec{r},t)$, the orbital counterpart to $\vec{n}(\vec{r},t)$, which allows us to incorporate the orbital dynamics into the system. Treating the orbital states as a pseudospin, we define $\vec{l}(\vec{r},t)$ as half the difference between the orbital pseudospins of the two sublattices [Fig.~\ref{fig:2}(c)]. Then, the extended Hamiltonian is given by 
\begin{equation}
	\mathcal{H} = \varepsilon_0 (\vec{k}) \tau_0 \otimes \sigma_0 + 
	\left[\vec{\tau} \cdot \vec{l}(\vec{r},t)  \right] \otimes \left[\varepsilon_\text{ani}(\vec{k}) \sigma_0+  J(\vec{k})\vec{\sigma} \cdot \vec{n}(\vec{r},t)\right],
	\label{3}
\end{equation}
where $\vec{\tau}$ and $\vec{\sigma}$ ($\tau_0$ and $\sigma_0$) denote the Pauli (identity) matrices in the orbital and spin spaces, respectively
. The rotational property of $J(\vec{k})$ reflects the nature of the spin splitting in altermagnets. For example, if $J\propto k_xk_y$, it corresponds to a $d$-wave magnet~\cite{roig2024minimal}. 
Within this framework, the explicit use of both $\vec{n}(\vec{r},t)$ and $\vec{l}(\vec{r},t)$ enables the Hamiltonian to describe the magnetic multipoles and their associated dynamics, as demonstrated below.

To examine the effects of spatiotemporal variations of the spin and orbital textures, we adopt a gauge transformation, which maps $\vec{n}(\vec{r},t)$ and $\vec{l}(\vec{r},t)$ to $n_z$ and $l_z$, respectively. 
For this purpose, we introduce unitary operators $U_s$ and $U_l$ acting on the spin and orbital spaces, defined as $U_s^\dagger = e^{i\theta\sigma_y/2} e^{i\phi\sigma_z/2}$ and $U_l^\dagger = e^{i\eta\tau_y/2} e^{i\chi\tau_z/2}$, respectively, where $\vec{n}(\vec{r},t)$ and $\vec{l}(\vec{r},t)$ in real space are expressed as $\vec{n} = (\sin\theta\cos\phi, \sin\theta\sin\phi, \cos\theta)$ [Fig.~\ref{fig:2}(a)] and $\vec{l} = (\sin\eta\cos\chi, \sin\eta\sin\chi, \cos\eta)$ [Fig.~\ref{fig:2}(c)] in spherical coordinate. The total unitary operator designed to diagonalize Eq.~\eqref{3} for both the spin and orbital parts is defined as $U_\text{tot}^\dagger = U_l^\dagger \otimes U_s^\dagger$, where $\otimes$ is the direct product of the orbital and spin spaces. As a result of the gauge transformation $\mathcal{H}'=U_\text{tot}^\dagger \mathcal{H} U_\text{tot} - i \hbar U_\text{tot}^\dagger \partial_t U_\text{tot}$, the orbital and spin states are determined by
$
\mathcal{H}'=\varepsilon_0 \left(\vec{\kappa}\right) \tau_0 \otimes \sigma_0 + 
\vec{\tau}_z \otimes \left[\varepsilon_\text{ani}(\vec{\kappa}) \sigma_0+  J(\vec{\kappa})\sigma_z \right] + e\mathbb{V}_\text{em}
$
, where $\vec{\kappa} = \vec{k}-e\mathbb{A}_\text{em}$. Here, the total vector potential operator $\mathbb{A}_\text{em}$ and scalar potential operator $\mathbb{V}_\text{em}$ are expressed as $\mathbb{A}_{\text{em}} = \mathbb{A}_l \otimes \sigma_0 + \tau_0 \otimes \mathbb{A}_s$ and $\mathbb{V}_{\text{em}} = \mathbb{V}_l \otimes \sigma_0 + \tau_0 \otimes \mathbb{V}_s$, where $\mathbb{A}_\nu = (i\hbar/e) U_\nu^\dagger \nabla U_\nu$ and $\mathbb{V}_\nu = -(i\hbar/e) U_\nu^\dagger \partial_t U_\nu$ for $\nu=l,s$.

The first-order correction due to the spatiotemporal variations results in the effective EEMFs felt by electrons under the adiabatic approximation~\cite{volovik1987linear,Faraday,nastt,Tatara2019}.  From the relations $\mathbb{E} = - \partial_t \mathbb{A} - \nabla \mathbb{V}$ and $\mathbb{B} = \nabla \times \mathbb{A}$, the $4 \times 4$ EEMF operators are derived as
\begin{align}
	\mathbb{E}_{\text{em}}= \mathbb{E}_l \otimes \sigma_0 + \tau_0 \otimes \mathbb{E}_s,~
	\mathbb{B}_{\text{em}}= \mathbb{B}_l \otimes \sigma_0 + \tau_0 \otimes \mathbb{B}_s.
	\label{5}
\end{align}
The EEMFs are then calculated by the statistical average of Eq.~\eqref{5} with respect to the distribution determined by the instantaneous Hamiltonian $\mathcal{H}'$. Since the zeroth-order term of $\mathcal{H}'$ is diagonal in the orbital and spin spaces, one can calculate the EEMFs by summing over the Fermi sea of the band-resolved diagonal elements of $\mathbb{E}_{\text{em}}$ and $\mathbb{B}_{\text{em}}$, which are written as
\begin{align}
	\vec{E}_\vec{d} 
	&= \frac{\hbar}{2e} \sum_i [\vec{d} \cdot (\partial_t \vec{d} \times \partial_i \vec{d})] \vhat{e}_i
	\label{6} \\
	\vec{B}_\vec{d} & = -\frac{\hbar}{4e} \sum_i \varepsilon^{ijk}[\vec{d} \cdot (\partial_j \vec{d} \times \partial_k \vec{d})] \vhat{e}_i,
	\label{7}
\end{align}
where $\vec{d} = \vec{n}, \vec{l}$ refers to both the spin and orbital N\'eel vectors.
There are several implications of our theory. First, although $\vec{n}$-driven EEMFs reproduce Eqs.~\eqref{1} and~\eqref{2}, the orbital N\'{e}el vector $\vec{l}$ generates a previously unrecognized physics. Second, the combination of these spin and orbital components forms distinct effective EEMFs across the four bands, which in turn drives the emergent transport phenomena of charge, spin, orbital and even magnetic multipoles. Third, even for charge and spin transport, the band anisotropy quantified by $\varepsilon_{\rm ani}(\vec{k})$ results in distinct physics absent in conventional antiferromagnets, as explicitly demonstrated in the next section.

\begin{figure}[t]
	\centering
	\includegraphics[width=1.0\linewidth]{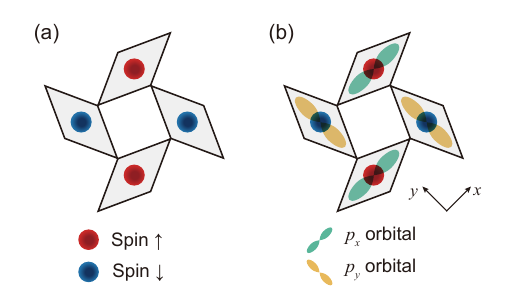}
	\caption{Schematic of lattice structure of the 2D d-wave altermagnet. (a) describes that the conventional Hamiltonian with spin degree of freedom and (b) indicates the modified Hamiltonian with orbital degree of freedom.}
	\label{fig:1}
\end{figure}

\section{Two-dimensional $d$-wave altermagnet}  
As an illustrative example, we consider a two-dimensional (2D) $d$-wave altermagnet, which has the lowest symmetry order and employs the $\{ \ket{p_x}, \ket{p_y} \}$ orbital basis. We assume that the $p_x$ orbital is dominant for the $\uparrow$-spin sublattice, and the $p_y$ orbital is dominant for $\downarrow$-spin sublattice (Fig.~\ref{fig:1}). 
To characterize the anisotropy of altermagnets 
we adopt an orbital-dependent effective masses, $m_1$ and $m_2$. In particular, $m_1$ describes the motion of $p_x (p_y)$ orbital electrons along the $x (y)$ direction, while $m_2$ governs the $p_x(p_y)$ electrons moving in the $y (x)$ direction, as inferred from the $\sigma$ and $\pi$ hoppings of $p$ orbitals. To separate the Hamiltonian into isotropic and anisotropic components, we define the reduced masses $\mu_1=m_1m_2/(m_1+m_2)$ and $\mu_2=m_1m_2/(m_2-m_1)$. 
By substituting $\varepsilon_0(\vec{k}) = \hbar^2 \vec{k}^2/2\mu_1$, $\varepsilon_{\text{ani}}(\vec{k}) = \hbar^2(k_x^2 - k_y^2)/2\mu_2$, and $J(\vec{k}) = J>0$ into Eq.~\eqref{3}, we obtain
\begin{equation}
	\mathcal{H} = \frac{\hbar^2\vec{k}^2}{2\mu_1} \tau_0 \otimes \sigma_0 + 
	\left[\vec{\tau} \cdot \vec{l}(\vec{r},t)  \right] \otimes \left[\frac{\hbar^2(k_x^2-k_y^2)}{2 \mu_2} \sigma_0+  J\vec{\sigma} \cdot \vec{n}(\vec{r},t)\right].
	\label{8}
\end{equation}
Equation~\eqref{8} is constructed in the basis of $\{\ket{p_x},\ket{p_y}\} \otimes \{ \ket{\uparrow},\ket{\downarrow}\}$, spanning the tensor-product orbital and spin Hilbert space. Assuming the strong-$J$ limit, one can effectively project Eq.~\eqref{8} onto $\{\ket{p_y}\otimes \ket{\uparrow},\ket{p_x}\otimes\ket{\uparrow}\}$ to obtain Eq.~\eqref{0} with $J_s(\vec{k})\propto (k_x^2-k_y^2)$. Furthermore, we have verified that the Eq.~\eqref{8} is invariant under the $[C_2||C_{4z}|t]$ spin space group of $d$-wave altermagnets~\cite{emerging,numerating} and has $\mathcal{PT}$ symmetry in the isotropic limit ($\mu_2 \rightarrow \infty$). 


\begin{subequations}\label{9}
To connect the band-resolved EEMFs to experimentally measurable quantities, we employ the Drude model, which successfully captures the essential transport behavior, as it is formally equivalent to the intraband Kubo formula within the relaxation time approximation with the relaxation time $\tau_\text{re}$. We note that the anomalous Hall effect in altermagnets may introduce further corrections to this Drude conductivity. First, we evaluate the longitudinal conductivity $\sigma_{ls,c}$, for each band using the Drude model, with the electron density of each band derived from Eq~\eqref{8}, where $l=\pm 1$ describes the $p_x$ and $p_y$ orbitals, respectively, and $s=\pm1$ describes the spin-$\uparrow$ and spin-$\downarrow$ states, respectively. The current of each band $\vec{j}_{ls}$ is driven by the emergent electric field via $\vec{j}_{ls}=\sigma_{ls,c} \vec{E}_{\text{em},ls}$. Combining these band contributions not only yields the charge $\vec{j}_e$ and spin $\vec{j}_s$ currents but also reveals the orbital current $\vec{j}_l$~\cite{orbital} and the magnetic octupole currents $\vec{j}_o = \sum_{ls} ls \vec{j}_{ls}$~\cite{octupole}. The calculated currents are
	\begin{gather}
		\vec{j}_e = 
		\sigma_- [\mathcal{M}_y(\vec{E}_\vec{l}-P\vec{E}_\vec{n})],\quad
		\vec{j}_s =  \sigma_+ [(-P\vec{E}_\vec{l}+\vec{E}_\vec{n})],
		\label{9-1} \\
		\vec{j}_l =  \sigma_+ [(\vec{E}_\vec{l}-P\vec{E}_\vec{n}) ],\quad
		\vec{j}_o=  
		\sigma_- [\mathcal{M}_y(-P\vec{E}_\vec{l}+\vec{E}_\vec{n}) ]
		\label{9-2},
	\end{gather}
where $\sigma_D = 2 e^2 \tau_{\mathrm{re}} \varepsilon_F / \pi \hbar^2$ is the 2D Drude conductivity, $\sigma_\pm = (\sigma_D /2) \left[\sqrt{(\mu_2 + \mu_1) / (\mu_2 - \mu_1) } \pm \sqrt{(\mu_2 - \mu_1) / (\mu_2 + \mu_1 )}\right]$, $\mathcal{M}_y$ is the operator that flips the sign of the $y$ component, such that $\mathcal{M}_y \mathbf{v} = (v_x, -v_y, v_z)$ for $v=(v_x,v_y,v_z)$, $P = J/\varepsilon_F$ is the polarization, and $\varepsilon_F$ denotes the Fermi energy. For more details, the band-resolved EEMFs and the conductivities for each band are summarized in Table~\ref{tab:table3}.
\end{subequations}

Notably, the dynamic orbital N\'{e}el vector $\vec{l}(\vec{r},t)$ not only contributes to $\vec{j}_l$ and $\vec{j}_o$ but also to $\vec{j}_e$ and $\vec{j}_s$. The emergent transport described by Eq.~\eqref{9} is fundamentally determined by band anisotropy. The coefficient $\sigma_+$ represents a quantitative anisotropic correction, whereas $\sigma_-$ is a purely anisotropy-driven contribution that reduces to zero in the isotropic limit ($\mu_2\to \infty$). This anisotropic nature is further manifested by the operator $\mathcal{M}_y$, which breaks rotational covariance. Consequently, the charge ($\vec{j}_e$) and magnetic octupole $(\vec{j}_o)$ currents strictly vanish in the isotropic limit, as their existence relies entirely on the non-zero $\sigma_-$ and the lattice anisotropy represented by $\mathcal{M}_y$. Regarding the physical implications, these results suggest that manipulating the anisotropy of an altermagnet enables not only the generation of charge and magnetic octupole currents but also the effective tuning of spin and orbital currents. Specifically, in our 2D $d$-wave altermagnet system, the Fermi energy can be shifted via an external gate voltage. This shift in the Fermi energy modifies the polarization, which in turn allows for precise control over the magnitudes of the various currents. This gate-dependent feature holds potential for transistor-like applications.

\begin{subequations}\label{Hall}
Next, we calculate the Hall conductivity, $\sigma_{ls,\text{H}}$, for each band using the Drude model. Unlike the longitudinal conductivity, the Hall conductivity is driven by the emergent magnetic field, where the 2D geometry ensures that only its $z$-component contributes to the transport. To investigate the first-order response to the EEMF, we evaluate the current of each band in the presence of an external electric field $\vec{E}_\text{ext} = E_\text{ext} \vhat{e}_x$.By summing these band contributions, the spin Hall $(\vec{j}_{s,\text{H}})_y$ and the orbital Hall $(\vec{j}_{l,\text{H}})_y$ current are given by
 \begin{align}
	(\vec{j}_{s,\text{H}})_y &= \frac{\sigma_D}{2} \left[ (1-P)F(\omega_{c+}) - (1+P)F(\omega_{c-}) \right] E_\text{ext}
	\label{10}, \\
	(\vec{j}_{l,\text{H}})_y &= \frac{\sigma_D}{2} \left[ (1-P)F(\omega_{c+}) + (1+P)F(\omega_{c-}) \right] E_\text{ext}
	\label{11},
\end{align}
where $F (\omega_{c\pm}) = \omega_{c\pm} \tau_\text{re} / [1+(\omega_{c\pm}\tau_\text{re})^2]$ and $\omega_{c \pm} = e[(\mathbf{B}_\mathbf{l})_z \pm (\mathbf{B}_\mathbf{n})_z]/\sqrt{\left(1/\mu_1\right)^2 - \left(1/\mu_2\right)^2}$ is the effective cyclotron frequency. These Hall currents occur in addition to the previously reported Hall contributions in altermagnets~\cite{Bai2023}. Similar to conventional antiferromagnets~\cite{TSHE0, TSHE1}, our results show that altermagnets do not exhibit the topological Hall effect due to the cancellation of charge current contributions from opposing sublattices, yet the topological spin Hall effect remains non-zero. While the topological spin Hall effect in conventional antiferromagnets is exclusively driven by spin-induced emergent magnetic fields, the topological spin Hall effect in altermagnets varies in magnitude depending on the magnetic anisotropy and is further influenced by the emergent magnetic fields generated by the orbitals.
\end{subequations}

These emergent magnetic fields generated by orbitals directly lead to the topological orbital Hall effect. Previous studies of the topological orbital Hall effect generally rely on spin-induced emergent fields, where spin textures such as skyrmions~\cite{Gobel2025Skyrmion} and hopfions~\cite{Gobel2025Hopfion} drive electrons into cyclotron motion, which in turn carry the orbital degree of freedom. In contrast, our work shows that the orbital N\'{e}el vector directly produces EEMFs. We also note the mutual influence, where conventional spin-induced fields affect the topological orbital Hall effect, and these orbital-derived EEMFs critically contribute to the topological spin Hall effect, as demonstrated by Eq.~\eqref{Hall} where each of $\vec{B}_\vec{n}$ and $\vec{B}_\vec{l}$ contributes to both $(\vec{j}_{s,H})_y$ and $(\vec{j}_{l,H})_y$. Consequently, the orbital-derived fields fundamentally govern the topological orbital Hall effect while simultaneously enhancing the topological spin Hall effect, establishing a novel mechanism where the orbital degree of freedom acts as the primary source of transport.

 
A fundamental prerequisite for the emergent transport discussed in this section is the existence of non-vanishing EEMFs. 
Various spin textures have been extensively investigated as potential sources of EEMFs. Theoretical studies have proposed a wide range of non-collinear configurations~\cite{Schulz2012}, including helical magnets driven by AC currents~\cite{Nagaosa2019}, domain walls and topological structures such as magnetic skyrmions~\cite{skyrmion1,skyrmion2}. On the experimental front, spin motive forces driven by emergent electric fields have been reported in canted helical states~\cite{Yokouchi2020,Kitaori2021} and vortex-type domain walls within Permalloy nanowires~\cite{SMFmeasure}, demonstrating the possibility of spintronic inductors. Just as domain wall motion generates a spin motive force in ferromagnets, recently proposed domain wall dynamics~\cite{Han2025Small} and STT~\cite{AMSTT,SST} in altermagnets suggest that altermagnetic textures can also drive a spin motive force. In addition, a possible generalization of $\vec{E}_\vec{n}$ by taking into account non-adiabaticity (see Appendix) broadens the possibility for realizing emergent electric fields.

On the other hand, real-space orbital textures have remained largely unexplored
. 
Previous studies have investigated the orbital Dzyaloshinskii-Moriya interaction~\cite{OrbitalDMI} and skyrmionic defects generated by vortex beams
~\cite{Vortexbeam1}. We propose that skyrmions of the orbital N\'{e}el vector (i.e., pseudospin) can be formed in altermagnets with strong orbital Dzyaloshinskii-Moriya interaction by locally irradiating spatially inhomogeneous light with circular polarization. The experimental realization of these textures remains a future challenge.

\begin{figure}[t]
	\centering
	\includegraphics[width=1.0\linewidth]{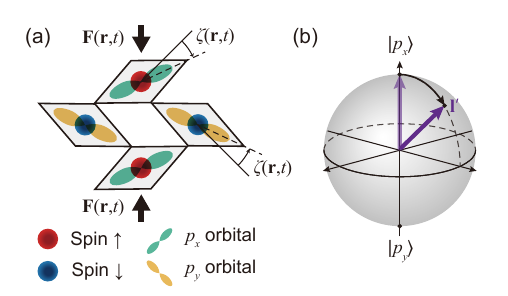}
	\caption{Configuration of the distorted lattice with distortion angle $\zeta$ for the 2D $d$-wave altermagnet. The lattice distortion is driven by external strain $F$.}
	\label{fig:3}
\end{figure}

\section{Orbital dynamics induced by strain variations}
An interesting feature of the extended model including the orbital degree of freedom is its ability to account for the roles of dynamic lattices~\cite{Han2025}.
As illustrated in Fig.~\ref{fig:3}, a strain may induce a lattice distortion that allows the orbitals to form a relative angle $\zeta(\vec{r},t)$, denoting the variation from the equilibrium state. In this case, the two orbital states become non-orthogonal to each other, and thus a novel feature may arise. Denoting the tilting angle by $\zeta(\vec{r},t)$, the orbital N\'{e}el vector $\vec{l}$ becomes $\vec{l}'=R(\vec{u};\pm \zeta)\vec{l}$, where $\vec{u}$ describes the rotational axis of $\vec{l}$ driven by strain. In this case, $\vec{u} = \vhat{e}_y$, which corresponds to orbital angular mixing. By substituting $\vec{l}'$ into Eqs.~\eqref{6} and~\eqref{7} and using the properties of rotational operators in the $\rm SO(3)$ group, the additional EEMFs are given by 
\begin{align}
	\vec{E}_\text{dis} &= \frac{\hbar}{2e} \sum_i \left[ (\partial_t \zeta) (\vec{u} \cdot \partial_i \vec{l}) - (\partial_i \zeta)(\vec{u} \cdot \partial_t \vec{l}) \right] \vhat{e}_i,
	\label{15} \\
	\vec{B}_\text{dis} &= - \frac{\hbar}{4e} \sum_i \varepsilon^{ijk} \left[ (\partial_j \zeta) (\vec{u} \cdot \partial_k \vec{l}) - (\partial_k \zeta)(\vec{u} \cdot \partial_j \vec{l}) \right] \vhat{e}_i.
	\label{16}
\end{align}
Equations~\eqref{15} and~\eqref{16} open another possibility to generate non-vanishing EEMFs in the presence of spatiotemporal variations in $\zeta(\vec{r},t)$. Both $\vec{E}_\text{dis}$ and $\vec{B}_\text{dis}$ are band-independent, consistent with their invariance under $\zeta \rightarrow -\zeta$ and $\vec{l} \rightarrow -\vec{l}$. Notably, the additional EEMFs are strictly orbital-dependent, which implies that the emergent transport driven by these fields is also purely orbital-dependent. 

\begin{figure}[t]
	\centering
	\includegraphics[width=1.0\linewidth]{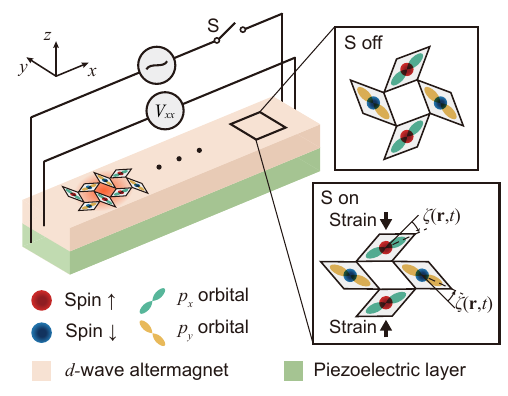}
	\caption{Schematic of the experimental setup to measure the charge current driven by the lattice-distortion-induced emergent electric field. Turning on the switch S applies an AC voltage to the green piezoelectric layer, inducing dynamic strain on the overlying $d$-wave altermagnet (peach). This strain drives the time-varying orbital Néel vector. An orange-shaded area represents circularly polarized light (illuminated area), which generates a real-space Néel vector texture.}
	\label{fig:4}
\end{figure}

As indicated by Eq.~\eqref{15}, generating an emergent electric field, $\vec{E}_\text{dis}$, requires the time dependence of the distortion angle ($\partial_t \zeta$) and the spatial gradient of the orbital texture ($\partial_i \vec{l}$). To experimentally realize this condition
, we propose the setup illustrated in Fig.~\ref{fig:4}. First, a 2D $d$-wave altermagnet is deposited onto a piezoelectric layer and patterned into a 1D nanowire along the $x$-axis. To provide the temporal variation ($\partial_t \zeta \neq 0$), the piezoelectric layer is driven by an AC current source, introducing a time-dependent strain. Simultaneously, local circularly polarized light (or vortex beams) may introduce a spatially varying orbital texture ($\partial_x \vec{l} \neq 0$). 
To isolate this signal from thermal artifacts such as the Seebeck effect, one can perform measurements for opposite circular polarizations and extract the voltage difference between the two cases.

\begin{subequations}\label{17}
As the band-dependent behaviors of $\vec{E}_\text{dis}$ and $\vec{B}_\text{dis}$ are different from the undistorted case, the corresponding current responses exhibit distinct features. After some algebra, the currents generated by $\vec{E}_\text{dis}$ are given by
\begin{align}
	\vec{j}_e^{\text{dis}} &=  \sigma_+ \vec{E}_\text{dis},\quad
	\vec{j}_s^{\text{dis}} = -\sigma_- P \mathcal{M}_y \vec{E}_\text{dis},
	\label{17-1} \\
	\vec{j}_l^{\text{dis}} &=\sigma_- \mathcal{M}_y \vec{E}_\text{dis},\quad
	\vec{j}_o^{\text{dis}} = -\sigma_+ P \vec{E}_\text{dis}.
	\label{17-2}
\end{align}
Note that the lattice-distortion-induced magnetic octupole current $\vec{j}_o^{\text{dis}}$ appears even in the isotropic limit ($\sigma_- \rightarrow 0$). Furthermore, the spin current $\vec{j}_s^{\text{dis}}$ and the magnetic octupole current $\vec{j}_o^{\text{dis}}$ are directly proportional to the polarization, suggesting that these emergent transport phenomena can be effectively tuned by an external gate voltage. In addition, lattice-distortion-induced orbital emergent magnetic field $\vec{B}_\text{dis}$ gives rise to the magnetic octupole Hall current, which vanishes in an undistorted lattice.
\end{subequations}

\section{Conclusion and Remarks}
In conclusion, we have extended the spin-based altermagnet Hamiltonian to include the dynamic orbital degree of freedom, allowing us to derive orbital-driven EEMFs
. We demonstrate 
the emergence of orbital and magnetic octupole transport, which are previously unreported, as well as anisotropic band contribution to the conventional charge and spin transport. 
Notably, the inclusion of the dynamic orbital degree of freedom allows us to examine the effects of spatiotemporal lattice distortions. We identify a purely orbital-dependent effect that drives additional magnetic octupole transport, which is absent in the undistorted case. We propose some experimental schemes to measure the proposed effects opening several related challenges and implying potentials for applications exploiting its gate-dependent behaviors.

Our study offers a new perspective on altermagnetism by incorporating the orbital degree of freedom, which provides a foundation for various directions in future research. Generalizing our approach to higher-order altermagnets~\cite{Beyond} will enable the study of the transport of higher-order magnetic multipoles, such as magnetic hexadecapoles and triakontadipoles.
For the dynamic lattice effect,
the Onsager reciprocal relation suggests that 
applying a charge, spin, orbital, or multipole current can generate a lattice wave, which is an active topic in recent research. Furthermore, while we considered a regime without spin-orbit coupling, introducing spin-orbit coupling would mix the spin and orbital spaces, potentially yielding a mixed-space Berry curvature~\cite{Hanke2017}.
Ultimately, we anticipate that our findings will stimulate further theoretical and experimental investigations deepening our knowledge of non-equilibrium orbital dynamics, as well as advancing both the fundamental understanding and technological potential of altermagnetic orbitronics.


\begin{acknowledgments}
\textit{Note added.---} During the preparation of our manuscript, we became aware of a recent theoretical work on emergent electrodynamics in altermagnets that focuses on the roles of spin textures~\cite{Schrade2026}.
	
We gretefully acknowledge D. Go, J. Kim, S. Kim, J. H. Han for cruical comments. This work was financially supported by the National Research Foundation of Korea (NRF) grant funded by the Korea government (MSIT) (RS-2024-00334933, RS-2024-00410027)
\end{acknowledgments}


\begin{thebibliography}{9}
\bibitem{octupole}
	The physical origin of the magnetic octupole current lies in the product of the orbital and spin indices, $ls$. In the $l=1$ subspace, the difference between the $p_x$ and $p_y$ orbital states is represented by the operator $L_x^2 - L_y^2$, which transforms spatially as $x^2 - y^2$. Coupling this orbital quadrupole character with the spin projection $S_z$ yields a term proportional to $(x^2 - y^2)S_z$, which corresponds to a magnetic octupole moment.
\end{thebibliography}

\newpage

\onecolumngrid
\EndMatter

\begin{table}[b]
	\renewcommand{\arraystretch}{3.0} 
	\begin{ruledtabular}
		\begin{tabular}{ccccc}
			\shortstack{Band \\ state} &
			\shortstack{Emergent electric \\ field ($\mathbf{E}_{\mathrm{em}}$)} &
			\shortstack{Emergent magnetic \\ field ($\mathbf{B}_{\mathrm{em}}$)} &
			\shortstack{Conductivity \\ ($\sigma_{ls,c}$)} &
			\shortstack{Hall conductivity \\ ($\sigma_{ls,\text{H}}$)}
			\\
			\colrule
			$(p_x,\uparrow)$ & $\mathbf{E}_\mathbf{l} + \mathbf{E}_\mathbf{n}$ & $\mathbf{B}_\mathbf{l} + \mathbf{B}_\mathbf{n}$ & 
			$\displaystyle \frac{(1-P)}{4} \begin{pmatrix} \sigma_+ + \sigma_- & 0 \\ 0 & \sigma_+ - \sigma_- \end{pmatrix}$ &
			$ \displaystyle \frac{(1-P)/4}{1+(\omega_{c+} \tau_{\mathrm{re}})^2} \begin{pmatrix} \sigma_+ + \sigma_- & - \omega_{c+} \tau_{\mathrm{re}} \sigma_D\\ \omega_{c+} \tau_{\mathrm{re}} \sigma_D & \sigma_+ - \sigma_- \end{pmatrix}$ \\
			
			$(p_x,\downarrow)$ & $\mathbf{E}_\mathbf{l} - \mathbf{E}_\mathbf{n}$ & $\mathbf{B}_\mathbf{l} - \mathbf{B}_\mathbf{n}$ & 
			$\displaystyle \frac{(1+P)}{4} \begin{pmatrix} \sigma_+ + \sigma_- & 0 \\ 0 & \sigma_+ - \sigma_- \end{pmatrix}$ &
			$\displaystyle \frac{(1+P)/4}{1+(\omega_{c-} \tau_{\mathrm{re}})^2} \begin{pmatrix} \sigma_+ + \sigma_- & - \omega_{c-} \tau_{\mathrm{re}} \\ \omega_{c-} \tau_{\mathrm{re}} & \sigma_+ - \sigma_- \end{pmatrix}$ \\
			
			$(p_y,\uparrow)$ & $- \mathbf{E}_\mathbf{l} + \mathbf{E}_\mathbf{n}$ & $- \mathbf{B}_\mathbf{l} + \mathbf{B}_\mathbf{n}$ & 
			$\displaystyle \frac{(1+P)}{4} \begin{pmatrix} \sigma_+ - \sigma_- & 0 \\ 0 & \sigma_+ + \sigma_- \end{pmatrix}$ &
			$\displaystyle \frac{(1+P)/4}{1+(\omega_{c-} \tau_{\mathrm{re}})^2} \begin{pmatrix} \sigma_+ - \sigma_- & \omega_{c-} \tau_{\mathrm{re}} \\ - \omega_{c-} \tau_{\mathrm{re}} & \sigma_+ + \sigma_- \end{pmatrix}$ \\
			
			$(p_y,\downarrow)$ & $- \mathbf{E}_\mathbf{l} - \mathbf{E}_\mathbf{n}$ & $- \mathbf{B}_\mathbf{l} - \mathbf{B}_\mathbf{n}$ & 
			$\displaystyle \frac{(1-P)}{4} \begin{pmatrix} \sigma_+ - \sigma_- & 0 \\ 0 & \sigma_+ + \sigma_- \end{pmatrix}$ &
			$\displaystyle \frac{(1-P)/4}{1+(\omega_{c+} \tau_{\mathrm{re}})^2} \begin{pmatrix} \sigma_+ - \sigma_- & \omega_{c+} \tau_{\mathrm{re}} \\ - \omega_{c+} \tau_{\mathrm{re}} & \sigma_+ + \sigma_- \end{pmatrix}$ \\
		\end{tabular}
	\end{ruledtabular}
	\caption{\label{tab:table3} EEMFs and corresponding conductivity tensors for each band of $d$-wave altermagnets. 
	\vspace{20pt}}
\end{table}
\twocolumngrid

\section*{Appendix A: Nonadiabatic contribution to $\rm \mathbf{E}_{\mathbf{n}}$} 
We began by formulating the Landau-Lifshitz-Gilbert equation to include nonadiabatic STT for each sublattice. By rewriting the expression in terms of the N\'{e}el vector $\vec{n}$ and magnetization $\vec{m}$, we used the Onsager reciprocal relation to determine the charge current for each band~\cite{nastt,Phenomology,Hals2010}. Consequently, the nonadiabatic STT-induced emergent electric field is given by
\begin{equation}
	\vec{E}_\text{NA} = s \beta \frac{g \mu_B}{2e^2N_s \gamma} \sum_i (\partial_t \vec{n} \cdot \partial_i \vec{n}) \vhat{e}_i,\label{12}
\end{equation}
where $\beta$ denotes the nonadiabatic STT term, $\gamma$ is a gyromagnetic ratio, $g$ is the electron $g$ factor, $N_s$ is a magnitude of N\'{e}el vector $\vec{n}$. Notably, the sign of Eq.~\eqref{12} is only depend on the direction of spin $s$. Furthermore, our calculation of the emergent electric field related to adiabatic STT term yields a result consistent with the spin contribution of $\vec{E}_\text{em}$ summarized in Table $\mathrm{I}$. Finally, the emergent charge and spin transport due to the nonadiabatic STT are driven by
\begin{equation}
	\vec{j}_e^{\text{NA}} = -\sigma_- P \mathcal{M}_y \vec{E}_\text{NA},~
	\vec{j}_s^{\text{NA}}  = \sigma_+ \vec{E}_\text{NA}.
\end{equation}
As it is not our primary interest, the contribution of $\beta$ to $\vec{B}_\vec{n}$ remains a future work. In addition, we suggest that the nonadiabatic effect of $\vec{E}_\vec{l}$ and $\vec{B}_\vec{l}$ would be interesting future challenges as the dynamics of orbitals in altermagnets are not completely understood.


\end{document}